\documentclass{PoS}

\usepackage{graphicx}
\usepackage{amsmath}
\usepackage{aas_macros}
\usepackage{xcolor}
\usepackage{accents}

\newcommand{\tmax}{t_{\rm max}}
\newcommand{\thetaj}{\theta_{\rm jet}}

\RequirePackage[left]{lineno}

\title{Preliminary Results of the \textit{Fermi} High-Latitude Extended Source Catalog}

\ShortTitle{Preliminary FHES Results}

\author{\speaker{Matthew Wood}\\
  Kavli Institute for Particle Astrophysics and Cosmology, SLAC National Accelerator Laboratory\\
  E-mail: \email{mdwood@slac.stanford.edu}}
\author{Jonathan Biteau\\
Insitut de Physique Nucl\'eaire d'Orsay, Universit\'e Paris-Sud, Univ. Paris/Saclay, CNRS/IN2P3, 91400 Orsay, France\\
E-mail: \email{biteau@ipno.in2p3.fr}
}
\author{Regina Caputo\\
  NASA, GSFC\\
  E-mail: \email{regina.caputo@nasa.gov}}
\author{Mattia Di Mauro\\
  Kavli Institute for Particle Astrophysics and Cosmology, SLAC National Accelerator Laboratory\\
  E-mail: \email{mdimauro@slac.stanford.edu}}
\author{Manuel Meyer\\
  Kavli Institute for Particle Astrophysics and Cosmology, SLAC National Accelerator Laboratory\\
  E-mail: \email{mameyer@stanford.edu}}

\author{on behalf of the \Fermi-LAT Collaboration}

\abstract{  

  We report on preliminary results from the \Fermi\ High-Latitude Extended Sources Catalog
  (FHES), a comprehensive search for spatially extended \gr\
  sources at high Galactic latitudes ($|b|>5\dg$) based on data from
  the \Fermi\ Large Area Telescope (LAT). While the majority of
  high-latitude LAT sources are extragalactic blazars that appear
  point-like within the LAT angular resolution, there are several
  physics scenarios that predict the existence of populations of
  spatially extended sources. If Dark Matter consists of Weakly
  Interacting Massive Particles, the annihilation or decay of these
  particles in subhalos of the Milky Way would appear as a population
  of unassociated \gr\ sources with finite angular extent.
  \gr\ emission from blazars could also be extended (so-called
  pair halos) due to the deflection of electron-positron pairs in the
  intergalactic magnetic field (IGMF). The pairs are produced in the
  absorption of gamma rays in the intergalactic medium and
  subsequently up-scatter photons of background radiation fields to
  \gr\ energies.  Measurement of pair halos could provide
  constraints on the strength and coherence length scale of the
  IGMF. 
  In a dedicated search, we find {\NUMEXTSOURCES} extended sources and
  {\NUMNEWEXTSOURCES} sources not previously characterized as
  extended. Limits on the flux of the extended source components are
  used to derive constraints on the strength of the IGMF using spectral and spatial templates derived
  from Monte Carlo simulations of electromagnetic cascades.  This
  allows us to constrain the IGMF to be stronger than
  $3\times10^{-16}\,$G for a coherence length
  $\lambda \gtrsim 10\,$kpc. }

\FullConference{35th International Cosmic Ray Conference -ICRC217-\\
		10-20 July, 2017\\
		Bexco, Busan, Korea}

\newcommand{\NUMROIS}{2693}

\newcommand{\NUMEXTSOURCES}{21}

\newcommand{\NUMKNOWNEXTSOURCES}{5}
\newcommand{\NUMNEWEXTSOURCES}{16}
\newcommand{\NUMNEWASSOCEXTSOURCES}{8}

\newcommand{\gr}{$\gamma$-ray}
\newcommand{\grs}{$\gamma$-rays}
\newcommand{\ee}{$e^+e^-$}

\newcommand{\liken}{\mathcal{L}_{\mathrm{n}}}

\newcommand{\likenplusm}{\mathcal{L}_{\mathrm{n} + \mathrm{m}}}

\newcommand{\tsext}{\mathrm{TS}_{\mathrm{ext}}}

\newcommand{\tshalo}{\mathrm{TS}_{\mathrm{halo}}}

\newlength{\dhatheight}

\newcommand{\fheshardiii}{FHES~J1741.5$-$3920}

\newcommand{\fhessofti}{FHES~J0000.9$+$6831}

\newcommand{\irf}[1]{\texttt{#1}}

\newcommand{\Figureref}[1]{Figure~\ref{fig:#1}}

\newcommand{\Fermi}{{\textit{Fermi}}}
\newcommand{\fermi}{\Fermi}

\usepackage{color}

\ifdefined\bwfigures
\else
\fi

\ifdefined \fdg \else
\newcommand\fdg{\mbox{$~.\!\!^\circ$}} 
\fi

\newcommand\dg{\mbox{$~\!\!^\circ$}}

\begin{document}


\section{Introduction}

The study of extended \gr\ sources gives insight into particle
acceleration, \gr\ emission processes, and enables searches for physics
beyond the Standard Model.  The predominant classes of spatially
extended \gr\ sources in the GeV band are supernova remnants (SNRs) and pulsar wind
nebulae (PWNe).  In these sources particle acceleration to \gr\
energies is believed to be driven by shocks in the expanding SNR shell
(in the case of SNRs) or at the termination shock of the pulsar wind
(in the case of PWNe).  A smaller class of extended \gr\ sources are
star-forming regions (SFRs) for which only a few examples are
currently known including the Cygnus Cocoon \cite{2011Sci...334.1103A}
and 30 Doradus region of the LMC
\cite{TheFermi-LAT:2015lxa,2015Sci...347..406H}.  Particle
acceleration in these objects may be driven by some combination of
stellar winds or successive supernova explosions.  Extended sources
are also found associated with the diffuse emission from nearby
galaxies: M31~\cite{Ackermann:2017nya}, Fornax~A
\cite{Ackermann:2016arn}, LMC \cite{TheFermi-LAT:2015lxa}, SMC
\cite{Fermi-LAT:2010fcp}, and Cen~A \cite{2010Sci...328..725A}.


Beyond the known classes of extended \gr\ sources, there are
several exotic scenarios that predict populations of extragalactic
extended sources.
If Dark Matter (DM) consists of Weakly Interacting Massive Particles
(WIMPs), the annihilation or decay of DM in subhalos of the Milky Way
could result in a population of spatially extended, unassociated,
\gr\ sources (e.g. \cite{Ackermann:2012nb}). 
Extended emission could also be associated with
the electromagnetic cascades generated by VHE gamma rays emitted by
\gr\ blazars.  The \ee~pairs produced when VHE photons
undergo pair production can in turn inverse-Compton (IC) scatter
photons of the cosmic microwave background (CMB), thereby initiating
the cascade~\cite{protheroe1993}.
The pairs are deflected in the intergalactic magnetic
field (IGMF) and, depending on its strength and coherence length, an
extended \gr\ halo can form around the AGN, often referred to as a pair
halo~\cite{aharonian1994}.  Measurement of pair halos could be used
to constrain the properties of the IGMF.

We report here on preliminary results of the Fermi High-Latitude Extended Sources Catalog
(FHES), a comprehensive search for spatially extended \gr\ sources
above 5\dg\ Galactic latitude using 7.5 years of \irf{Pass 8} data
above 1~GeV.  We further examine the spatial and spectral properties
of the TeV~blazars in our sample to look for evidence of pair halos
and derive constraints on the IGMF.

\section{Data Analysis}

We analyze 90 months of LAT data (2008 August 4 to 2016 February 4)
selecting \irf{P8R2 SOURCE}-class events in the energy range from
1~GeV to 1~TeV.  We perform a binned maximum-likelihood analysis with
eight logarithmic bins per decade in energy and a region of interest
(ROI) of $6\dg \times 6\dg$ with an angular pixelization of
$0\fdg025$.  To maximize the sensitivity to small angular
extensions, we split the data into two independent samples
according to the quality of the angular reconstruction:
\irf{evtype}$=$32 (PSF3) and \irf{evtype}$=$28 (PSF0+PSF1+PSF2), where PSF stands for point spread function.  

We consider \NUMROIS~regions of interest (ROIs) centered on sources
listed in the 3FGL and 3FHL catalogs \cite{3fgl,2017arXiv170200664T}
with $|b| > 5\dg$.  We use the Galactic Interstellar Emission Model
(IEM) recommended for \irf{Pass8} analysis (\texttt{gll\_iem\_v06.fits})
and standard templates for the isotropic background.  We include in
our model 3FGL sources within a region of $10\dg \times10 \dg$
centered on each ROI.  We exclude 3FGL sources detected with a Test
Statistic (TS)\footnote{TS is defined as twice the log likelihood
  ratio between models with and without the source.} less than 100 or
analysis flags indicating confusion with diffuse emission (flags 5, 6,
or 8).

We optimize each ROI by first re-fitting the spectral parameters of
all model components and the positions of point sources inside the ROI
boundary.  We then identify new point source candidates in two passes.
We first identify new point sources in the outer ROI ($R > 1.0\dg$)
with $\mathrm{TS} > 9$ and fit the spectrum and position for each.  We
then iteratively search for new point sources in the inner ROI while
testing the central source for extension.  With each additional point
source we test the hypothesis of $n$ point sources against two
alternative hypotheses: substituting the central source with a 2D
Gaussian with the same spectral parameterization (extension
hypothesis) and superimposing the central source with a 2D Gaussian at
the same position with a power-law spectrum (halo hypothesis).  We
stop adding new point sources when no additional candidates are found
or when a model of extension is found to be preferred according to the
Akaike information criterion (AIC, see Ref.~\cite{Akaike}) given by
$\mathrm{AIC} = 2k - 2\ln\mathcal{L}$ where $k$ is the number of
parameters in the model.

After selecting the best-fit iteration $n$, we evaluate the
statistical evidence for extended emission from the likelihood ratio
of models with and without an extended component,

\begin{equation}\label{eqn:tsext}
\mathrm{TS}_\mathrm{m} = 2\left(\ln\likenplusm - \ln\liken\right),
\end{equation}
where $\mathrm{m} =$ ext (halo) for the extension (halo) hypothesis,
$\liken$ is the likelihood for the model with $n$ point sources, and
$\likenplusm$ is the likelihood for the halo or extension hypothesis.
We identify a source as extended if
$\tsext>16$ 
or $\tshalo > 16$ where in the latter case we reanalyze the source
after adding an extended component with the spatial and spectral
parameters of the best-fit halo.


\section{Extended Source Catalog}

\begin{figure}[t]
\centering
\includegraphics[width=0.99\columnwidth]{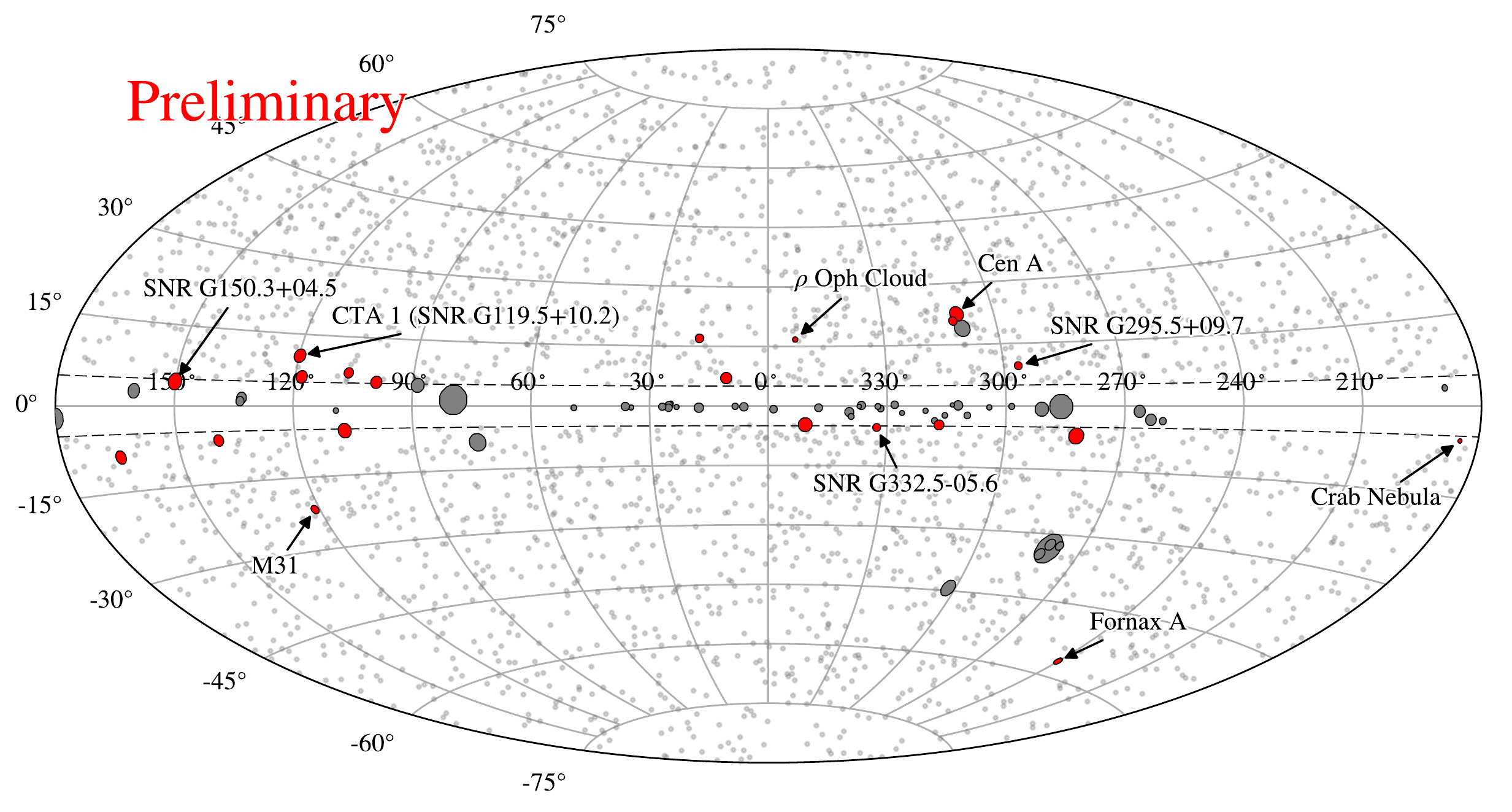}
\caption{Distribution of FHES sources in Galactic coordinates.  Light
  grey markers indicate sources that are best fit by a
  point-source morphology ($\tsext < 16$).  Red circles with black
  outline indicate FHES sources that are best fit by an extended
  morphology ($\tsext > 16$).  The size of the marker is drawn to
  scale of the intrinsic 68\% containment radius of the source.
  Labeled sources are those with a previously published detection of
  extension or an unambiguous association to a multiwavelength
  counterpart. Grey circles with black outline indicate the position
  and angular size of known LAT extended sources that fell outside our
  latitude selection or were explicitly excluded from the
  analysis. \label{fig:results_allsky} }
\end{figure}

Our analysis detects {\NUMEXTSOURCES} sources with statistically significant
evidence for extension ($\tsext > 16$).
\Figureref{results_allsky} shows the the distribution of the new
extended source candidates.  Of these extended sources,
{\NUMKNOWNEXTSOURCES} were previously detected as extended and
{\NUMNEWEXTSOURCES} are new.  Of the new sources,
{\NUMNEWASSOCEXTSOURCES} have potential associations including two
sources associated to Cen~A, two SNRs, and an extended source consistent with the position of the Crab Nebula.  
However, the significance of the extension of this source drops significantly and is below $\mathrm{TS}_\mathrm{ext} = 16$ if systematic uncertainties on the PSF are considered.
The remaining eight unassociated sources all have
$|b| < 20\dg$ consistent with a Galactic origin.



\begin{figure}[t]
\centering
\includegraphics[width=0.49\columnwidth]{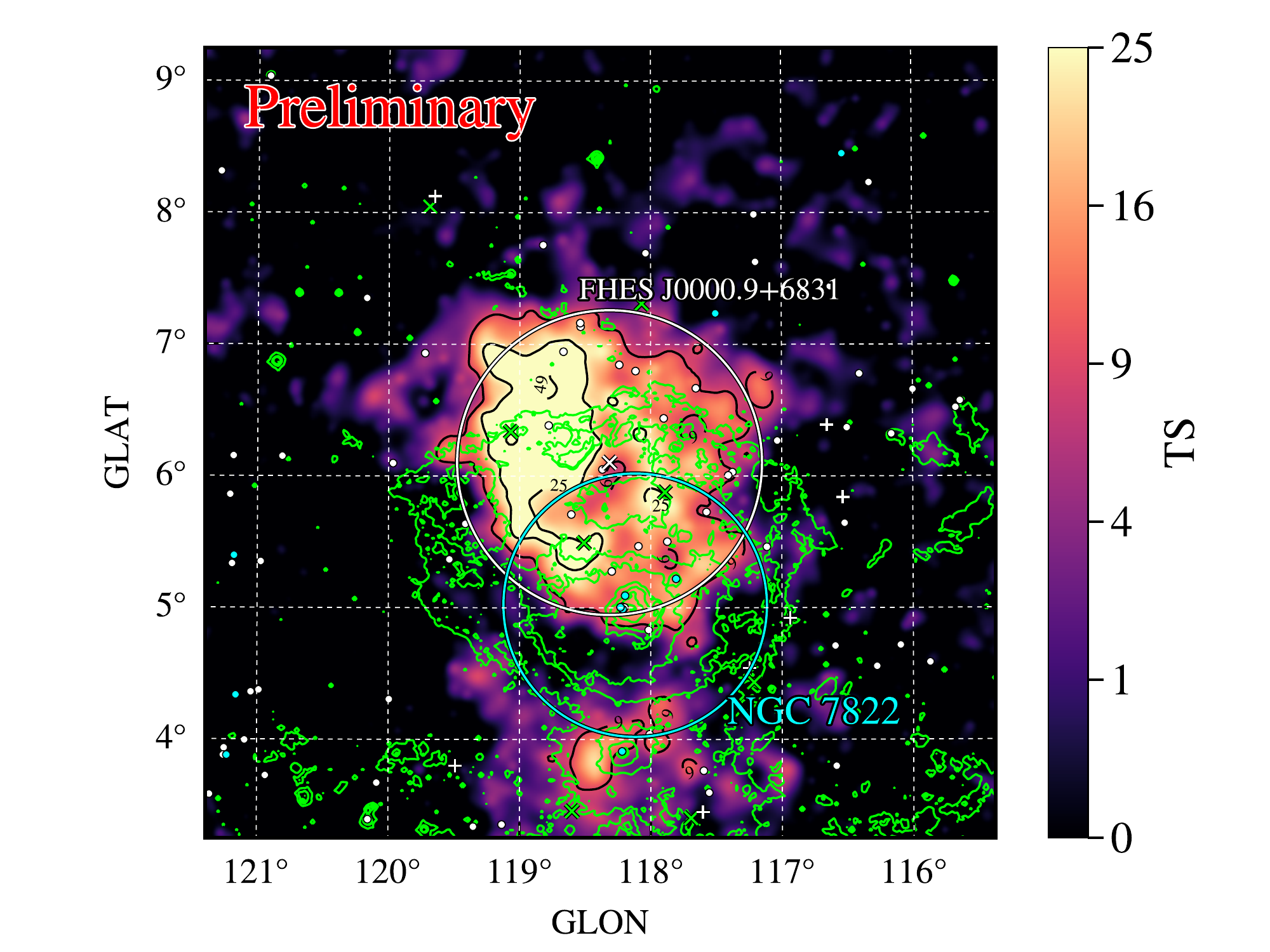}
\includegraphics[width=0.49\columnwidth]{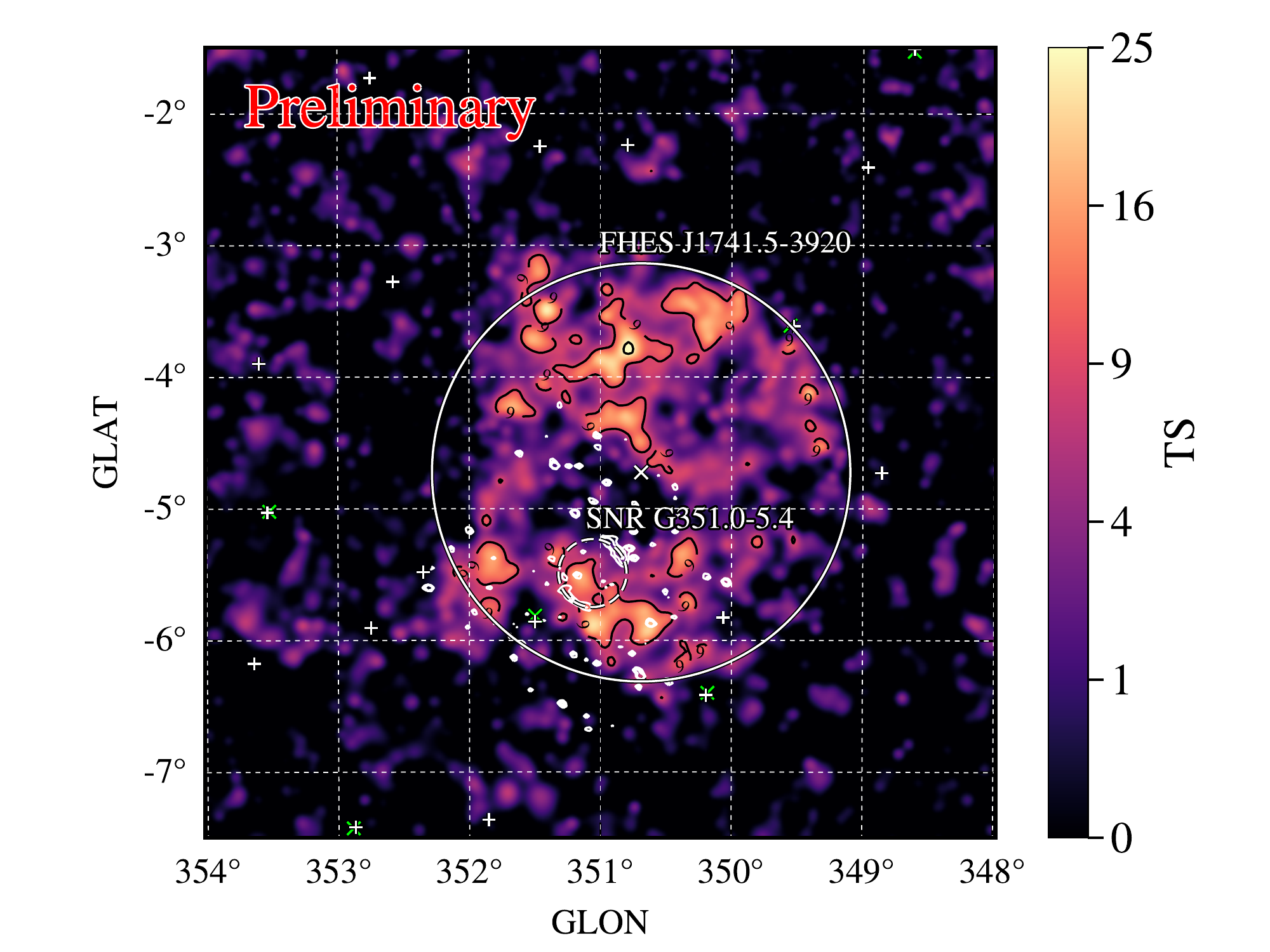}
\caption{Residual TS maps for the FHES extended source candidates
  {\fhessofti} (left) and {\fheshardiii} (right). The white circle
  with central cross indicates the best-fit disk extension and
  centroid of the FHES source.  White pluses indicate point-source
  candidates from the best-fit model for the region.  Green crosses
  indicate the positions of sources in the 3FGL catalog.  White and
  cyans circles indicate the locations of O and B stars from the
  SIMBAD database.  Green contours in the left plot show the WISE map
  of 22~$\mu$m emission. 
  The cyan circle indicates the extent of the HII region NGC~7822
  which is tentatively associated to {\fhessofti}.  White contours and
  white dashed circle in the right panel show the GMRT radio map of
  SNR~G351.0$-$5.4 at 325~MHz \cite{deGasperin:2014caa} and its
  inferred position and diameter.\label{fig:tsmaps} }
\end{figure}

We identify three sources that we tentatively associate to SFRs by
their spatial correlation with multiwavelength tracers of
star-formation activity.  All three sources are found to be robust to
IEM systematics and have no apparent correlation to features in the
gas or dust maps of the respective regions.  The left panel of
\Figureref{tsmaps} shows the residual TS map for {\fhessofti}, an
extended FHES sources that we associate to the star-forming HII region
NGC~7822.  {\fhessofti} has an extension of $0.95\dg$ and a soft-\gr\
spectrum ($\Gamma\simeq2.7$) that is consistent with the typical
spectrum arising from CR interactions with the ISM.  NGC~7822 is part
of the Cepheus OB4 star-forming region which is located at a distance
of 1~kpc.  An overdensity of O and B stars is observed within the
region encompassed by {\fhessofti} indicating that the \gr\ source may
be associated with the collision of stellar winds driven by O/B stars.
A correlation is also observed with a shell-like feature in the map of
emission at 22~$\mu$m from WISE which traces the thermal emission of
dust heated by young, massive stars in the region.

An example of a source without clear association is {\fheshardiii},
shown in the right panel of \Figureref{tsmaps}.  It has the largest
angular size ($D\sim3.0\dg$) and hardest \gr\ spectrum
($\Gamma = 1.8 \pm 0.04$) of the unassociated FHES sources.  We note
that \cite{Araya:2016cbn} report the discovery of a new \gr\ source
with the same position, morphology, and spectral characteristics.  The
spectrum and morphology of this source suggest that it may be
associated with a young shell-type remnant.  {\fheshardiii} is
positionally coincident with
SNR~G351.0$-$5.4~\cite{deGasperin:2014caa}, a radio-detected SNR with
$D\sim0.5\dg$.  However, an association with this SNR is disfavored
given the large mismatch in angular size.

\section{IGMF Limits}

We do not find evidence for  an extension of 
blazars and use the FHES to
derive constraints on the coherence length, $\lambda$, and field
strength, $B$, of the IGMF.  We use both spectral and spatial
information of the catalog as well as spectra from imaging air
Cherenkov telescopes (IACTs) to derive these constraints.  For this
study we select 9 BL~Lac objects with known redshifts and
well-measured TeV spectra.

We use the ELMAG Monte Carlo code \cite{2012CoPhC.183.1036K} to
generate models for the cascade component associated with each object
in our sample.  This code computes the expected flux and angular
distribution of the cascade by propagating individual \grs\ and
modeling the electromagnetic cascades produced when these photons pair
convert on CMB and EBL photons.  ELMAG uses a simplified model of the
IGMF that splits the field into cells with constant field strength and
size equal to the coherence length scale.  In addition to the IGMF
parameters, the cascade depends on the intrinsic properties of the
blazar: the redshift ($z$), the spectrum ($\phi(\mathbf{p})$) with
intrinsic parameters $\mathbf{p}$, the jet opening angle ($\thetaj$),
and the source activity time scale ($\tmax$).  We simulate the full
cascade spectrum over a grid of redshifts and in bins of injected \gr\
energy $\Delta E$ between 100\,MeV and 32\,TeV (using again 8 bins per
decade) using the EBL model of~\cite{Dominguez:2010bv}.  Using these
simulations we compute
a grid of models for the flux and angular size of the cascade as a
function of the IGMF and source parameters.  We derive limits on the
IGMF by performing a joint fit to the LAT and IACT measurements of
each blazar.  We find the best-fit IGMF parameters while
simultaneously fitting the parameters of the intrinsic spectrum
($\mathbf{p}$).


%
%
\begin{figure*}[t]
\centering
\includegraphics[width = .49\linewidth]{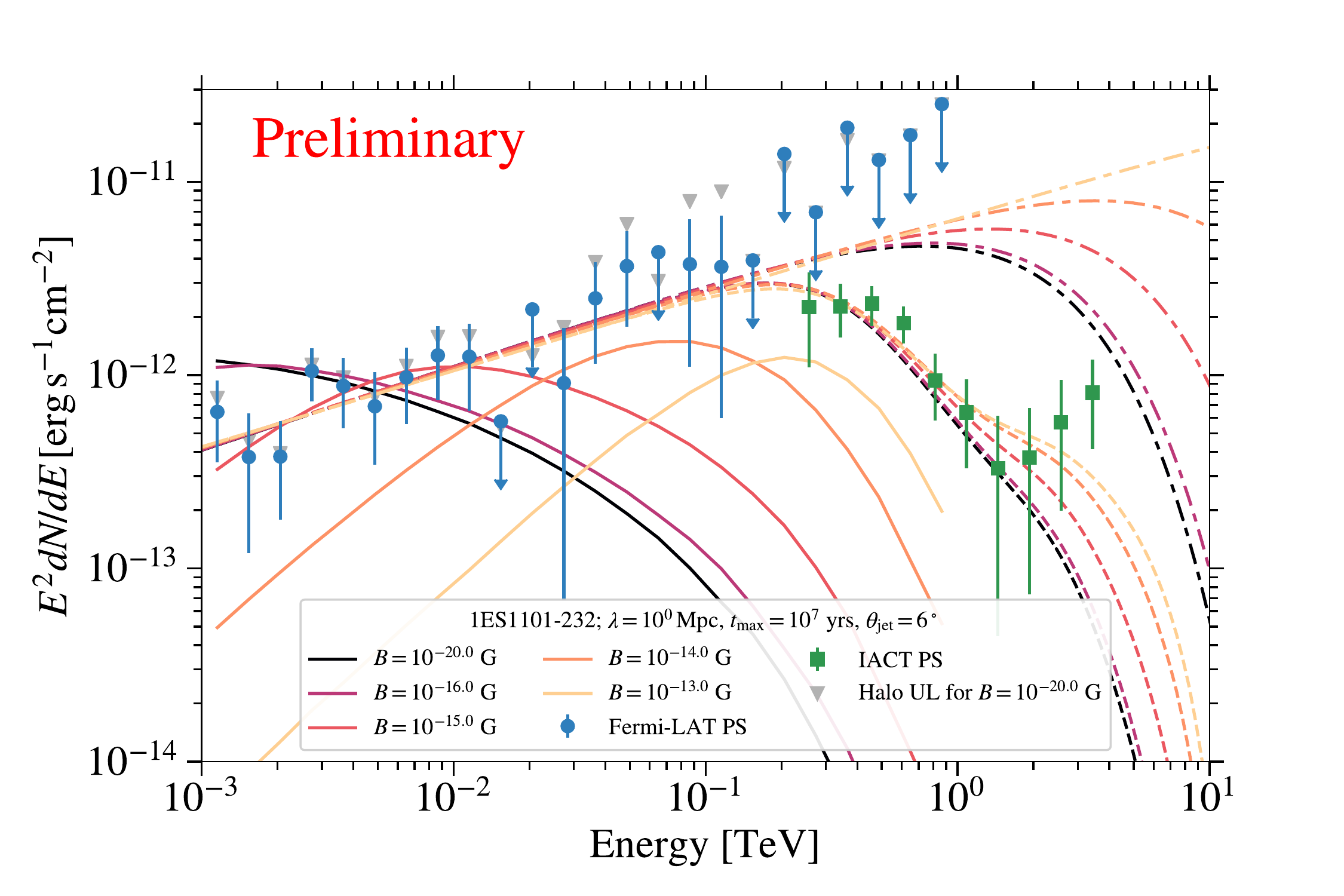}
\includegraphics[width = .49\linewidth]{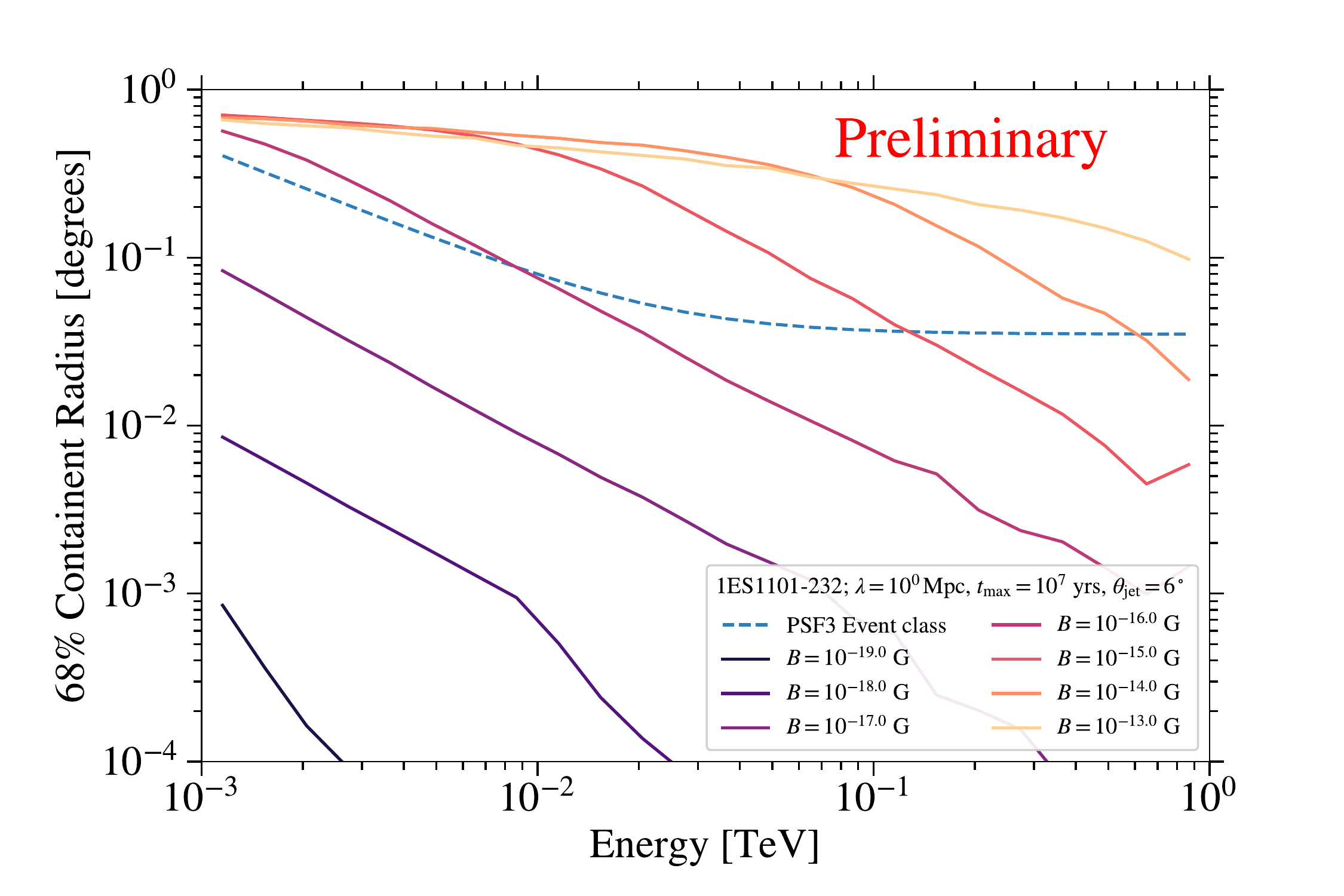}
\caption{\textit{Left}: Fit of the intrinsic spectrum and cascade
  component to the IACT and \fermi-LAT data of 1ES\,1101-232
  ($z = 0.186$) for different IGMF strengths. A source activity time
  of $t_\mathrm{max} = 10^7$\,years and a jet opening angle of
  $\theta_{\rm jet} = 6^\circ$ are assumed together with a coherence
  length of 1\,Mpc. The best-fit intrinsic spectra multiplied with EBL
  absorption are shown as dashed lines with colors matching those of
  the cascade component (solid lines). The intrinsic spectra are shown
  as dotted lines. Upper limits on the halo energy flux for widths
  equal to that of the cascade for $B = 10^{-20}$\,G are shown as gray
  triangles. \textit{Right}: Containment radii for the cascade
  for different $B$-field strengths and the PSF (PSF3 event
  type) as a function of energy for the same source and parameters as
  the right panel.
\label{fig:sed-igmf}
}
\end{figure*}
%
%

The source parameters $\thetaj$ and $\tmax$ are poorly constrained for
the sources under consideration.  For $\thetaj$ we choose $6\dg$ which
is consistent with the typical blazar jet angles measured from radio
observations.  For $\tmax$ we consider timescales
$t_{\rm max} = 10,\ 10^4,\ 10^7$\,years where $\tmax = 10$~yr is chosen
to match the length of the LAT observations and $\tmax = 10^7$~yr 
is the expected activity timescale of AGN~\cite{parma2002}.


The left panel of Figure~\ref{fig:sed-igmf} shows the best-fit
spectrum and cascade contribution versus magnetic field strength for
one of the objects in our samples, 1ES\,1101-232.  The right panel of
Figure~\ref{fig:sed-igmf} shows the angular size of the cascade as a
function of energy and magnetic field strength.  In the limit of small
magnetic field strengths the cascade component peaks at GeV energies
and has an angular size well below the LAT angular resolution.  For
larger field strengths the peak of the cascade moves to higher
energies.  The angular extent of the halo is measurable when
$B \gtrsim 10^{-16}$\,G.

\begin{figure*}[t]
\centering
\includegraphics[width = 0.49\linewidth]{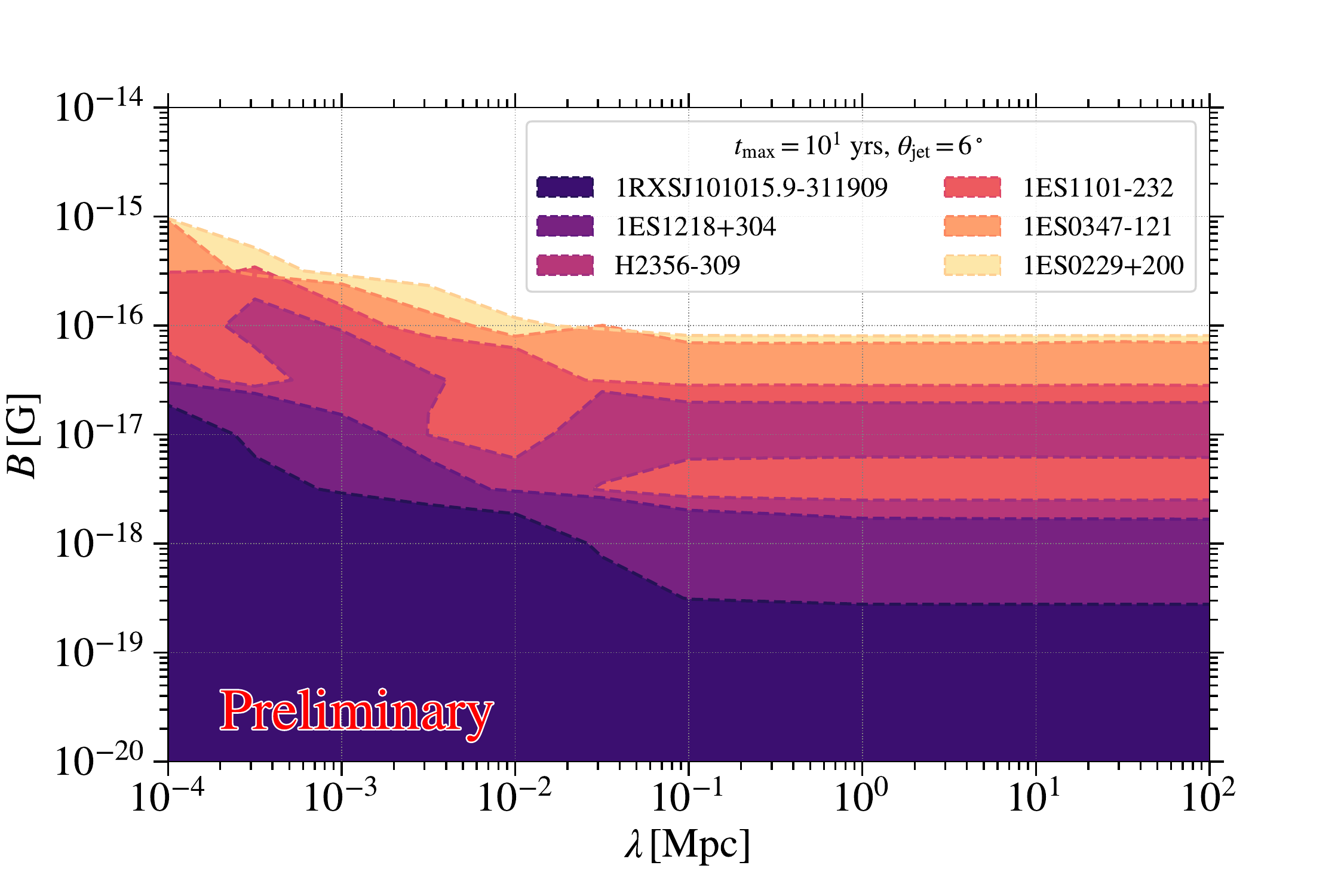}
\includegraphics[width = 0.49\linewidth]{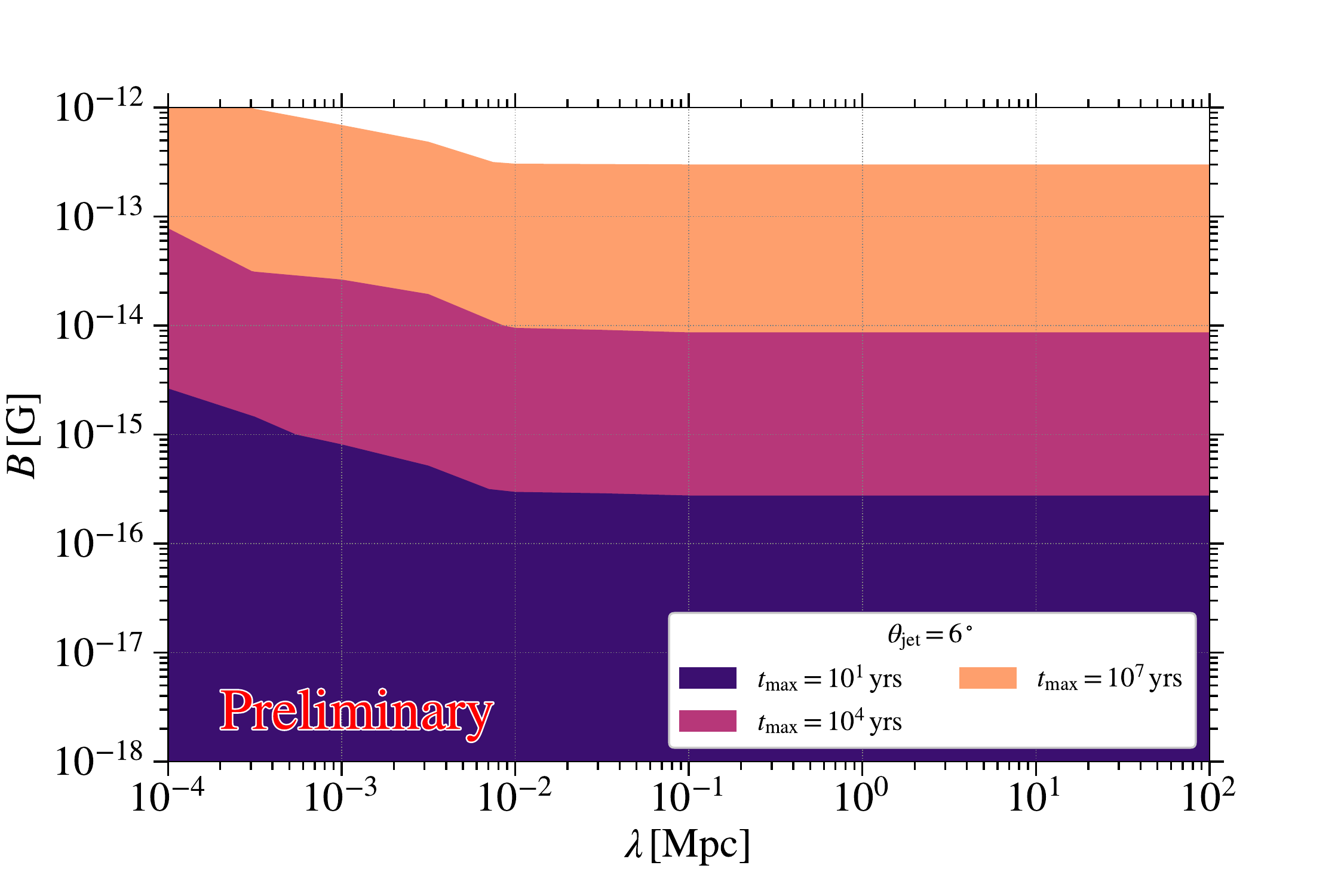}
\caption{\label{fig:like} 95\,\% lower limits on the field strength of
  the IGMF for $\theta_{\rm jet} = 6^\circ$. \textit{Left:} Exclusions
  for $t_{\rm max} = 10\,$years for individual
  sources. \textit{Right:} Combined exclusion limits for different
  blazar activity times.  
  }
\end{figure*}

We find in all cases that the best-fit model is one with a large
$B$-field that suppresses the cascade flux to a level below what would
be detectable in the LAT energy regime.  We obtain 95\% limits in the
IGMF parameter space by finding the values of $(B,\lambda)$ that
change $2\ln\mathcal{L}$ by 5.99 corresponding to a $\chi^2$
distribution with two degrees of freedom.  \Figureref{like} shows the
IGMF limits obtained for individual blazars as well as combined limits
from stacking the likelihoods of all of the BL~Lac objects in the IGMF
sample.  Under the short activity timescale scenario
($\tmax = 10$~yr), we are able to constrain
$B \gtrsim 3\times10^{-16}$\,G for $\lambda > 10^{-2}$\,Mpc.  For the
long timescale scenario ($\tmax = 10^{7}$~yr), we can constrain
$B \gtrsim 3\times10^{-13}\,$G for $\lambda > 10^{-2}$\,Mpc.

\section{Conclusions}

We report on preliminary results of a search for extended sources at
high Galactic latitude.  We are able to identify {\NUMEXTSOURCES}
extended sources, {\NUMNEWEXTSOURCES} of which are identified as such
for the first time.  All of the unassociated sources are found with
$|b| < 20\dg$ indicating that they have either a Galactic origin or
arise from systematic uncertainties in the IEM.  We do not find
evidence for extended emission associated with extragalactic source
populations either from blazars or unassociated sources.

Using the results of the extended source catalog, we are able to
derive constraints on the IGMF that are substantially stronger than
previous works using LAT observations of TeV blazars (see
e.g. \cite{finke2015}), limiting $B \gtrsim 3\times10^{-16}$\,G for
$\lambda \gtrsim 10\,$~kpc for an activity time of the considered
blazars of 10\,years.  Using instead activity times of $10^4$ ($10^7$)
years improves the limits to $B \gtrsim 9\times10^{-15}$\,G
($B \gtrsim 3\times10^{-13}$\,G).  For such large fields, however, the
actual jet opening and viewing angle of the blazar become important to
accurately model the halo.  The influence of these effects in the
limit of large field strength ($B \gtrsim 10^{-15}$\,G) is not
considered in the simplified 1D Monte-Carlo calculation used by ELMAG.

Dedicated 3D Monte Carlo codes should be used in the future to search
for the cascade emission at higher values of the IGMF to accurately
model the source extension and taking into account the viewing angle
of the
blazar~\cite{2010ApJ...719L.130N,2016PhRvD..94h3005A,2017MNRAS.466.3472F}.
Further extensions could include more realistic models of the
intergalactic field, including a full treatment of its turbulence
spectrum~\cite{caprini2015} and its helicity~\cite{chen2015helical}.


\acknowledgments

The \textit{Fermi}-LAT Collaboration acknowledges support for LAT
development, operation and data analysis from NASA and DOE (United
States), CEA/Irfu and IN2P3/CNRS (France), ASI and INFN (Italy), MEXT,
KEK, and JAXA (Japan), and the K.A.~Wallenberg Foundation, the Swedish
Research Council and the National Space Board (Sweden). Science
analysis support in the operations phase from INAF (Italy) and CNES
(France) is also gratefully acknowledged.

\bibliographystyle{JHEP}
\bibliography{../extension_main}

\end{document}